\documentclass[12pt]{iopart}
\usepackage{amssymb}
\usepackage{graphicx}

\newcommand{\beq}{\begin{equation}}
\newcommand{\eeq}{\end{equation}}

\begin{document}

\title[Efficient QKD secure against no-signalling eavesdroppers]{Efficient quantum key distribution secure against no-signalling eavesdroppers}

\author{Antonio Ac\'\i n$^1$, Serge Massar$^2$, Stefano Pironio$^1$}
\address{$^1$ICFO-Institute de Ciencies Fotoniques, Mediterranean Technology Park, 08860 Castelldefels (Barcelona), Spain}
\address{$^2$Laboratoire d'Information Quantique and Centre for Quantum Information and Communication, {C.P.} 165/59, Universit\'{e} Libre de Bruxelles, Avenue F. D. Roosevelt 50, 1050 Bruxelles, Belgium}
\eads{\mailto{Antonio.Acin@icfo.es}, \mailto{smassar@ulb.ac.be}, \mailto{Stefano.Pironio@icfo.es}}

\begin{abstract}
By carrying out measurements on entangled states, two parties can generate a secret key which is secure not only against an eavesdropper bound by the laws of quantum mechanics, but also against a hypothetical ``post-quantum'' eavesdroppers limited by the no-signalling principle only. We introduce a family of quantum key distribution protocols of this type, which are more efficient than previous ones, both in terms of key rate and noise resistance. Interestingly, the best protocols involve large number of measurements. We show that in the absence of noise, these protocols can yield one secret bit per entanglement bit, implying that the key rates in the no-signalling post-quantum scenario are comparable to the key rates in usual quantum key distribution.
\end{abstract}
\pacs{0.365.Ud, 0.367.Dd, 0.367-a}

\section{Introduction}
A quantum key distribution (QKD) protocol allows two parties sharing a quantum communication channel to exchange a secret key for later cryptographic purposes. The hypothesis which ensures the security of QKD is that an eavesdropper trying to acquire knowledge about the key is bound by the laws of quantum mechanics \cite{BB84,reviewQKD}. Recently, it has been realized that the non-local correlations of entangled states can be exploited to make QKD secure against an eavesdropper that is limited by the no-signalling principle alone. That is, the only assumption made on the eavesdropper is that it cannot prepare two or more physical systems in a joint state such that a local measurement on one system may transfer information to another, distinct system. Compatibility with special relativity justifies this assumption in the case that the systems are spacelike separated. That the principle of no-signalling alone is sufficient to guarantee the security of a QKD scheme was demonstrated in \cite{BHK}. A practical scheme, tolerating in particular a finite amount of experimental noise and producing a non-zero key rate, was proposed in \cite{AG}, and a similar protocol was outlined in \cite{bhk} (these latter works, however, do not take into account the most general type of attacks available to an eavesdropper).

In the protocol of \cite{AG}, Alice and Bob share maximally entangled states and perform on each pair of particles measurements maximizing the violation of the Clauser-Horne-Shimony-Holt (CHSH) inequality \cite{chsh}. The resulting measurement outcomes are used to establish the secret key.

In this paper, we combine the ideas of \cite{AG} with those of the Ekert scheme \cite{Ekert} and those of \cite{lca}. We introduce a protocol where the parties carry out the CHSH non-locality test on a subset of their particles only, but otherwise perform their measurements in the same basis. The purpose of the CHSH test is to guarantee that the eavesdropper Eve has limited knowledge about Alice and Bob's system, as follows from the monogamy property of non-local correlations \cite{blmppr,bkp}. The highly correlated outcomes obtained when Alice and Bob measure in the same basis are used to establish the secret key, thereby maximizing the key production rate. In this way, we obtain a protocol that has a key rate
higher than the one presented in \cite{AG}, and which is also significantly more noise resistant.

Further improvements are obtained by using for the security test the chained inequalities \cite{pearle,bc} for $N$ measurement settings instead of the CHSH inequality.
In the absence of noise these protocols have increasing key rates for larger $N$. As $N\to \infty$ it becomes possible to extract one bit of secret key per e-bit shared by Alice and Bob. This is a consequence of a property noted in \cite{BHK,bkp}: that in the limit $N\to \infty$, the correlations  maximally violating the chained inequality are fully monogamous, in the sense that Eve cannot get any information about the measurement outcomes of Alice and Bob. This shows that the key rates that can be extracted against eavesdroppers limited by quantum mechanics, or limited by no-signalling only, are similar.

In this work, as in \cite{AG}, we restrict our security analysis to individual attacks, that is, to an eavesdropper trying to acquire knowledge about Alice and Bob's individual systems independently and always in the same way. Further comments on the possibility of proving security against more general attacks are given in the conclusion.

\section{Protocol based on the CHSH test}
We begin by presenting the protocol based on the CHSH inequality in detail, before extending the approach to the chained inequalities.
Alice and Bob share a quantum channel consisting of a source that emits pairs of qubits in the maximally entangled state
\begin{equation}\label{ebit}
|\phi_+\rangle = \left(|0\rangle_A|0\rangle_B + |1\rangle_A |1\rangle_B \right)/\sqrt{2}\,.
\end{equation}
As usual in quantum key distribution, we consider that imperfections are present. For definiteness, we will assume that the effect of the noise is to transform the state~\eref{ebit} in the Werner state
\begin{equation}\label{werner}
\rho = p\, |\phi_+\rangle \langle \phi_+ |  + (1-p) \frac{I}{4}\,,
\end{equation}
and we will study the performance of the protocol as a function of $p$. But our analysis can easily be extended to other states and other forms of noise, as it does not depend on the specific form of the quantum state shared by Alice and Bob or of the measurements performed, but only on the probability distribution characterizing the results of  these measurements.

On each of their qubits, Alice and Bob perform measurements chosen randomly and independently. We denote Alice's measurements by $x$, and those of Bob by $y$, and the corresponding outcomes $a$ and $b$. The joint probability to obtain outcomes $a$ and $b$ given measurement $x$ and $y$ will be denoted $P(ab|xy)$.

Alice has a choice between three measurements $x=0$, $1$, or $2$, corresponding to measuring her qubit in the bases $|0\rangle\pm e^{i\phi(x)}|1\rangle$,  where $\phi(0)=\pi/4$, $\phi(1)=0$ and $\phi(2)=\pi/2$. The probability that she chooses the measurement along $\pi/4$ is $q$, and the probabilities that she measures along $0$ or $\pi/2$ are both $(1-q)/2$. Bob has a choice between two measurements $y=0$ or $1$, corresponding to the bases $|0\rangle\pm e^{-i\phi(y)}|1\rangle$, with $\phi(0)=\pi/4$ and $\phi(1)=-\pi/4$. The probability that he measures along $\pi/4$ is $q'$ and along $-\pi/4$ is $1-q'$.

After all the pair of particles have been measured, Alice and Bob reveal their measurement basis. If Alice measured along $0$ or $\pi/2$, they are in the situation maximizing the violation of the CHSH inequality. In this case, they both reveal the result of their measurements. From these results, they compute the expectation value of the CHSH expression.
If Alice measured along $\pi/4$ and Bob measured along $-\pi/4$, the measurement results are completely uncorrelated. The data corresponding to these cases is thrown away.
If they both measured along $\pi/4$, the measurement outcomes are strongly correlated and will serve as a raw key. In a later step, they will carry out information reconciliation and privacy amplification to obtain a pair of identical secret keys from this raw data.

As we show below, the security of the protocol follows from the combined facts that Alice and Bob observe a violation of the CHSH inequality and that Eve is bound by the no-signalling condition. We note anticipatively that the security does not depend on the probabilities $q$ and $q'$ with which they choose the measurements $x=0$ and $y=0$. This is because the purpose of the other measurements is to verify that they share the correct conditional probabilities $P(ab|xy)$ of obtaining the outcomes $a$ and $b$ given the measurements $x$ and $y$. It is these conditional probabilities that are constrained by the no-signalling conditions. The probabilities with which Alice and Bob choose $x$ and $y$ are irrelevant, so long as they accumulate sufficient statistics to determine $P(ab|xy)$ with precision. They will therefore choose $q$ and $q'$ as close as possible to 1 to maximize the key generation rate, leaving only a few instances where other measurements are made.

\section{Eavesdropping strategies}
We assume that the source of particles is situated between Alice and Bob and is under the control of the eavesdropper Eve. Eve has thus the ability to prepare the particles of Alice and Bob and any physical system in her possession in a joint, no-signalling (possibly non-quantum) state. As mentioned earlier, we restrict our analysis to individual attacks, i.e., to attacks where Eve acquires independent knowledge about each individual bit of the key. This amounts to consider that for each pair, Eve prepares a state of three particles, one for each of Alice, Bob, and herself\footnote{While in the most general type of attacks, Eve could prepare a single, collective state describing her system and the $2n$ particles sent to Alice and Bob.}.
This state defines the measurement probabilities $P(abe|xyz)$, where $z$ denotes a possible measurement performed by Eve on her system, and $e$ the resulting outcome. The only constraint imposed on this joint distribution is that it is no-signalling, i.e., that it satisfies the conditions
\begin{equation}
\sum_e P(abe|xyz)\equiv P(ab|xy) \quad\mbox{for all } z\,,
\end{equation}
and the analogous ones obtained by summing over Alice's and Bob's outputs. These conditions imply that the marginal distributions for any subset of the parties are independent of the choices of measurements made by the other parties.

The objective for Eve is to perform a measurement on her particle that will give her maximal information about Alice's and Bob's outcomes. As usual when considering individual attacks, we assume that Eve performs her measurement before the information reconciliation and privacy amplification phase. Note that Eve is in fact only interested by Alice's and Bob's results for the pair of observables ($x=0,y=0$), since the outcomes for the other pairs of measurements are not part of the key, and are announced anyway. Moreover, by no-signalling, Eve cannot influence these outcomes by her choice of measurement. We conclude that there is no point for her to vary her measurement depending on which observables are measured by Alice and Bob, and we can therefore assume that she always performs the same measurement $\tilde z$, the one that gives her optimal information for the pair ($x=0,y=0$). Since she does the same measurement every time, she may as well do it right away after the particles have been created. Conditioned on getting the output $e$, which happens with probability $p_e=P(e|\tilde z)$, the net effect of Eve's strategy is thus to prepare Alice and Bob's particles in a state characterized by the no-signalling probabilities $P_e(ab|xy)=P(ab|xy\tilde z e)$.

We have thus reduced individual attacks by a no-signalling eavesdropper to the preparation of a mixture $\sum_e p_e\, P_e(ab|xy)$ of no-signalling distributions, which should obviously return the observed correlations between Alice and Bob: $P(ab|xy)=\sum_e p_e\, P_e(ab|xy)$. In this decomposition, Eve's knowledge is represented by the variable~$e$.

Clearly, the most general strategy for Eve corresponds to a mixture where each no-signalling term $P_e(ab|xy)$ is \emph{extremal}, i.e., cannot itself be written as a convex sum of other no-signalling correlations. In our protocol, $a$ and $b$ take binary values. The extremal points of the set of no-signalling correlations in this case have been described in \cite{BP,JM}. For each measurement $x$ of Alice, they are two possibilities. Either the output $a$ of $x$ is predetermined, i.e., $P(a|x)=0,1$; or it is uniformly random, i.e., $P(a|x)=1/2$. In this latter case, the measurement $x$ is part of a set of two measurements for Alice and a set of two measurements for Bob such that restricted to this set, the CHSH inequality is violated up to its algebraic maximum. The situation is similar for Bob.

\section{Security analysis}

Having described the strategy of Eve, we now analyze the security of our protocol. Eve is constrained by the measured values of the distribution $P(ab|xy)$, which in turn determine the degree of violation of the CHSH inequality and the amount of correlations between Alice and Bob when they measure in the bases $x=0$ and $y=0$. For the Werner state \eref{werner} and the measurements of our protocol, the average value of the CHSH expression is
\beq\eqalign{\label{chsh}
\langle \mbox{\textit{CHSH}}\rangle &= P(a_1\neq b_0)+P(a_1\neq b_1)
+ P(a_2\neq b_1) + P(a_2=b_0)\\
&=2-\sqrt{2}\,p\,.}
\eeq
Note that with this notation for the CHSH expression, local correlations satisfy $\langle \mbox{\textit{CHSH}}\rangle\geq 1$, and general non-local ones $\langle \mbox{\textit{CHSH}}\rangle\geq 0$.
The correlations between Alice and Bob when they measure the pair $(x=0,y=0)$ can be quantified by the quantity
\beq\label{corr}
\langle C \rangle = P(a_0=b_0)-P(a_0\neq b_0)= p\,.
\eeq

Using the results of \cite{BP,JM} mentioned above, the extremal strategies available to Eve can be classified in different sets, according to whether the measurements $x=0$ or $y=0$ yield predetermined outcomes or yield uniformly random outcomes. Since Bob has a choice between two measurements only, the option that $y=0$ yields a deterministic outcome implies in fact that all measurements yield deterministic outcomes. We can thus group Eve's strategies in three sets, corresponding to the cases where $x=0$ and $y=0$ both yield deterministic outcomes, where $x=0$ yields a deterministic outcome and $y=0$ yields a locally random outcome, and where $x=0$ and $y=0$ both yield locally random outcomes. For each set, we can set a bound on $\langle \mbox{\textit{CHSH}}\rangle$ and $\langle C \rangle$, and we can also determine the conditional entropies $H(A|E)$ and $H(B|E)$, representing the ignorance of Eve on the raw key, and the conditional mutual information $I(A:B|E)$. These properties are summarized in Table \ref{tab1}.
\begin{table}[h]
\caption{\label{tab1}Extremal strategies available to Eve. The first line corresponds to strategies where the measurements $x=0$ and $y=0$ used to establish the secret key both yield deterministic outcomes $(D,D)$, the second to the case where the measurement $x=0$ yields a deterministic outcome and the measurement $y=0$ yields a locally random outcome $(D,R)$, and the third line to the situation where both outcomes are locally random $(R,R)$.}
\begin{indented}\item[]
        \begin{tabular}{ccccccc}\br
        & Strategies & $\langle \mbox{\textit{CHSH}}\rangle$ & $\langle C\rangle$ & $H(A|E)$ & $H(B|E)$ & $I(A:B|E)$\\
\mr
    1&$(D,D)$& $\geq 1$ &
$\leq 1$ &
$0$&
$0$&
$0$\\
2&$(D,R)$& $\geq 0$ &
$0$ &
$0$&
$1$&
$0$\\
3&$(R,R)$& $\geq 0$ &
$\leq 1$ &
$1$&
$1$&
$1$\\
\br
\end{tabular}
\end{indented}
\end{table}

We can now apply the Csisz\'{a}r-K\"{o}rner condition \cite{CK}, which gives the secret key rate for privacy amplification with one-way communication: $K=\max\{I(A:B)-I(A:E),I(A:B)-I(B:E)\}$.
From Table \ref{tab1} we note that there is an asymmetry because Eve's ignorance on Bob's outcome is larger than on Alice's outcome in case 2, implying that the mutual information $I(A:E)$ between Alice and Eve will be larger than the mutual information $I(B:E)$ between Bob and Eve. Hence to optimize the key rate the communication during the privacy amplification protocol should go from Bob to Alice, with the corresponding rate given by $K=I(A:B)-I(B:E)$.

The mutual information between Alice and Bob is $I(A:B)=1-h(1/2+p/2)$, where $h$ is the binary entropy. To compute the mutual information between Bob and Eve, we denote $p_1$, $p_2$, $p_3$ the probability that Eve chooses strategy $1$, $2$, $3$ above. Then, $I(B:E)=H(B)-\sum_i p_i\, H_i(B|E)$. Using the values of the conditional entropies given in Table \ref{tab1} and $H(B)=1$, we find $I(B:E)=p_1$. This probability is constrained by the fact that the $p_i$ are positive, sum to one, and yield the measured values of $\langle \mbox{\textit{CHSH}}\rangle$ and $\langle C\rangle$. These last two conditions take the form
\beq\label{pchsh}
p_1\leq \langle \mbox{\textit{CHSH}} \rangle
\eeq
and
\beq\label{pcorr}
p_1+p_3\geq \langle C \rangle\,.
\eeq
Using condition \eref{pchsh} and the value \eref{chsh}, we
deduce that the key rate is related to the violation of the CHSH inequality through
\beq \label{KK}
K\geq \sqrt{2}p-h(\frac{1+p}{2})-1\,,
\eeq
with equality attained if Eve saturates the inequalities of Table \ref{tab1}.
In the absence of noise ($p=1$), we find $K\geq\sqrt{2}-1\simeq 0.414$ which is almost 4 times larger than the noise-free rate of the protocol described in \cite{AG}.
The key rate vanishes ($K=0$) when $p=0.9038$ whereas the key rate for the protocol described in \cite{AG} vanishes when $p=0.9319$. The present protocol is thus also more noise resistant.

It is also interesting to compute a bound on the intrinsic
information $I_{\downarrow}=\linebreak[4]I(A:B{\downarrow} E)$ which is an upper bound on the key rate using two-way key distillation protocols \cite{mwo}. The intrinsic information is defined as $I(A:B{\downarrow}E)=\linebreak[4]\min_{E\rightarrow \bar{E}}\{I(A:B|\bar{E}\}$, where the minimization is taken over all channels mapping Eve's random variables $e$ onto new random variables $\bar e$ with probabilities $P_{\bar e|e}$. From this definition, it follows that $I_\downarrow$ is upper bounded by the conditional mutual information, $I(A:B{\downarrow}E)\leq I(A:B|E)$. This allows us to write $I_\downarrow\leq \sum_i p_i\, I_i(A:B|E)=p_3$. Using first \Eref{pcorr} and then \Eref{pchsh} with inequalities replaced by equalities (thus assuming that Eve uses the optimal strategy saturating the inequalities of Table \ref{tab1}), we obtain
$I_\downarrow\leq \langle C \rangle- \langle \mbox{\textit{CHSH}} \rangle$. Inserting the values \eref{chsh} and \eref{corr}, we find $I_\downarrow \leq (1 + \sqrt{2})p -2$. We deduce that the intrinsic information vanishes, and hence the protocol becomes useless, when $p= 2 / (1 + \sqrt{2})\simeq 0.8284$. Note, however, that the state \eref{werner} violates the CHSH inequality for $p\geq 0.7071$, and that a set of correlations violating a Bell inequality (before any further processing) always has a positive intrinsic information \cite{mag}. Whether a key can be extracted from such data, and if it can, what is the best protocol achieving it, remains an open question.

Finally, we mention that the security of our protocol can also be analyzed against a standard quantum eavesdropper. Under this assumption, our protocol is equivalent to the BB84 protocol and has the same key-rate vs noise performance. The same remark applies to the family of protocols introduced below.

\section{Generalizations of the protocol}
Our protocol can be generalized in a systematic way. Consider
some non-locality test and the associated Bell inequality $I$. On a fraction of their particles, Alice and Bob will carry out the non-locality test and determine the amount of violation of $I$. This constrains Eve to use non-local strategies, for which she has only limited information about Alice and Bob's outcomes. In order to generate the secret key, Alice and Bob will perform measurements that yield strongly correlated outcomes (e.g., if they share singlets, they perform their measurements in the same basis), such that at least one of Alice's or Bob's measurements is also used in the non-locality test. The performance of this key distribution scheme depends on the properties of the Bell inequality: what fraction of times Eve can fix Alice's or Bob's outcomes, how resistant to noise is the Bell inequality, etc. We illustrate this
general approach by introducing a family of protocols based on the chained Bell inequality for $N$ measurements, an inequality already used in \cite{BHK}. The protocol corresponding to $N=2$ coincides with the one we just presented.

This family of protocols is defined as follows. Alice carry out $N+1$ measurements $x=0,\dots,N$ corresponding to measurements in the bases
$\{ |0\rangle \pm e^{i \phi(x)} |1\rangle\}$, with $\phi(0)=\pi/2N$ and $\phi(x)=\pi x/N$ for $x=1,\ldots, N$. Bob has a choice between $N$ measurements $y=0,\ldots N-1$ corresponding to measuring in the bases $\{|0\rangle \pm e^{-i \phi(y)} |1\rangle\}$, where $\phi(y)=\pi (y+1/2)/N$.

As before, most of the time Alice and Bob choose to measure $x=0$ and $y=0$, and obtain highly correlated bits which serve as the raw data used to establish the secret key.  The mutual information between Alice and Bob is thus $I(A:B)=1-h(1/2+p/2)$. The other measurements are used to determine $P(ab|xy)$ and in particular to check the violation of the chained inequality:
\beq\label{chain}
\langle \mbox{\textit{CHAIN}}\rangle=\sum_{i=1}^{N} \left[P(a_{i}\neq b_{i-1})+P(a_{i}\neq b_{i})\right]\,,
\eeq
where $b_N$ stands for $b_0+1\mbox{ mod }2$. With the measurements mentioned above, we find $\langle \mbox{\textit{CHAIN}}\rangle=2N\sin^2(\pi/4N)$ for the maximally entangled state \eref{ebit}, and
\beq\label{chain2}
\langle \mbox{\textit{CHAIN}}\rangle=N(1-p\,\cos(\pi/2N))
\eeq
if Alice and Bob share the Werner state \eref{werner}.

\begin{figure}[h]
\begin{indented}\item[]
\includegraphics[scale=0.74]{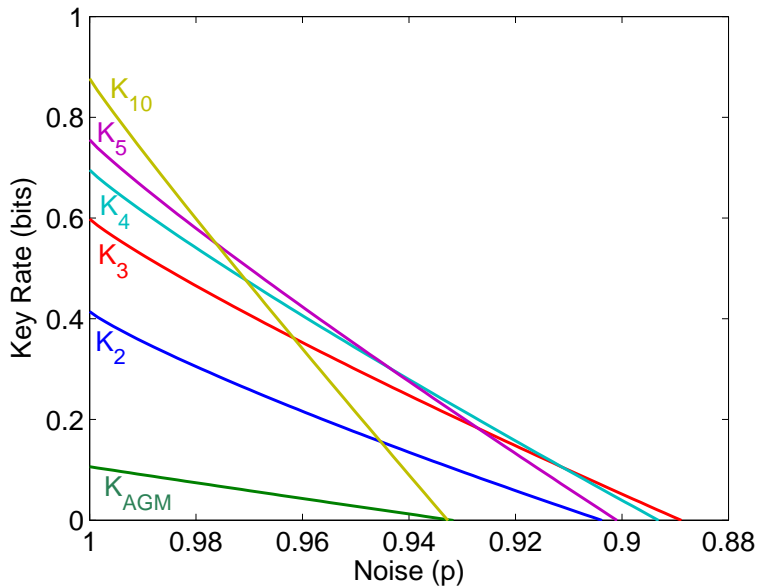}
\caption{Key rate for different protocols versus amount of noise (represented by the purity $p$ of the Werner state \eref{werner}). $K_{AGM}$ denotes the key rate for the protocol of \cite{AG}. $K_N$, given by (\ref{KN}),  represents the key rate for the family of protocols studied here.
Note that $K_2$ is uniformly better than $K_{AGM}$, that $K_3$ is uniformly better than $K_2$, and that for large $N$ (illustrated by $N=10$), and in the absence of noise ($p=0$), the key rate tends to $1$.
}
\label{fig:epsart}\end{indented}
\end{figure}

Let us now consider the strategies available to an eavesdropper limited only by no-signalling. As before note first that because the measurement $x=0$ is not part of the non-locality test, Eve can fix completely the corresponding output and at the same time produce an arbitrary violation of the chained inequality. Eve's knowledge on Alice's outcome will thus generally be larger than on Bob's outcome, and therefore the communication in the privacy amplification phase should go from Bob to Alice. For each no-signalling extremal distribution, there are two possibilities: either the input $y=0$ has a deterministic outcome or it has a completely random outcome. It was shown in \cite{bkp}, that for any measurement $y$ used in the chained inequality, the following bound holds: $\langle \mbox{\textit{CHAIN}}\rangle \geq 2\,P(b|y)-1$. When $y=0$ has a deterministic outcome, we therefore deduce that $\langle \mbox{\textit{CHAIN}}\rangle\geq 1$, and we also have $H(B|E)=0$. When $y=0$ has a uniformly random outcome, $\langle \mbox{\textit{CHAIN}}\rangle\geq 0$ and $H(B|E)=1$. From these properties, it follows that
\beq
I(B:E)\leq \langle \mbox{\textit{CHAIN}}\rangle\,,
\eeq
and hence that the key rate using one-way privacy amplification is lower bounded by
\beq\eqalign{
K_N&\geq 1-h(\frac{1+p}{2})-N(1- p\,\cos(\frac{\pi}{2N}))
\label{KN}
\\
&\gtrsim 1-h(\frac{1+p}{2}) -p\,\frac{\pi^2}{8N}-N(1-p) \ .}\eeq
The protocols corresponding to $N=3,4,5$ are more efficient than the CHSH based protocol for all noise levels, as illustrated in Figure \ref{fig:epsart}.
The best overall noise resistance is achieved for $N=3$, and corresponds to $p=0.8889$. As $N$ increases the protocols
become increasingly sensitive to noise, since they require $p\geq 1-O(1/N)$.
But in the ideal case where there is no noise, the key rate tends to one, $K_N \geq 1 - \pi^2/8 N$. This follows from a property noted in \cite{bkp}: that in the absence of noise, and when $N=\infty$, the correlations that we introduced are maximally non-local in the sense that they do not admit any local component. Hence Eve must always use non-local strategies, for which she has zero knowledge about Bob's outcome.

\section{Variant of our protocol with preprocessing of the raw key}
\begin{figure}[h]
\begin{indented}\item[]
\includegraphics[scale=0.74]{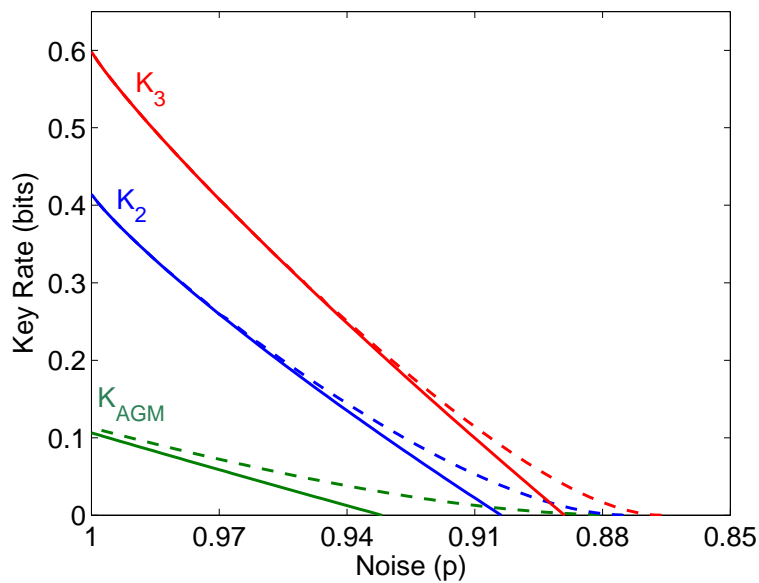}
\caption{Effect of the preprocessing discussed in Sec.~6 on the key rate. Solid and dotted lines correspond, respectively, to protocols without and with preprocessing. $K_2$ and $K_{AGM}$ vanish for $p=0.8740$. The best overall noise resistance is obtained for $N=3$ and corresponds to $p=0.8660$.
}
\label{fig:epsart2}\end{indented}
\end{figure}

Following the ideas suggested in \cite{kgr}, we introduce in this section a variant of our protocol where the parties perform a preprocessing of the raw key before carrying out the information reconciliation and privacy amplification stage. This preprocessing consists of Bob flipping his outcome with probability $r$ and leaving it unchanged with probability $1-r$. Although this operation disturbs the perfect correlations between Alice and Bob, it also prevents Eve to have complete information on Bob's outcome, even when she uses a local deterministic strategy. The effect of this preprocessing step on the efficiency of the protocol is illustrated in Figure \ref{fig:epsart2}, where for each value of $p$, we have numerically determined the optimal value of $r$. The result of the pre-processing is to increase the key rate, particularly for small values of $p$, and to improve the noise resistance.

Note that with preprocessing, the protocol of \cite{AG} and our protocol with $N=2$ have both a key rate that vanishes for $p=0.8740$. The best noise resistance for all $N$ is obtained as before for $N=3$, with a positive key rate for $p\geq 0.8660$.

\section{Discussion and conclusion}
We have analyzed key distribution against individual attacks by an eavesdropper limited
only by no-signalling. The detection of nonlocal correlations is a
necessary condition for security in this scenario \cite{BHK,AG}.
But beyond this qualitative assertion, little is known on how to
optimally exploit the nonlocal correlations of quantum mechanics to establish a
secret key. We have shown here that by adding one measurement to the CHSH test, the key rate and the noise resistance can significantly be improved. The idea is to use nonlocal
correlations only for estimating Eve's knowledge, but to build the key from perfectly correlated measurements. The resulting protocol is very close in spirit to Ekert's protocol, but with the tools at
our disposal fifteen years after Ekert's seminal work, we understand
much better the origin of the security.

We have argued that this approach can be based on any non-locality
test, and as an illustration have studied a family of protocols
based on the chained inequalities for $N$ measurement settings. Each inequality in the family provides a
different estimation of Eve's knowledge. When $N$ is large, the corresponding protocols are very sensitive to noise. But in the absence of noise, they allow Alice and Bob to extract asymptotically one secret bit per e-bit. If the noise is important, it becomes advantageous to use a chained inequality with fewer measurements to put a strong bound on Eve's knowledge. In general, to maximize the key rate Alice and Bob should estimate the properties of
their channel and adapt the non-locality test to the measured
parameters. Using the chained inequalities as non-locality tests, the optimal key rate is then the interpolation of
all the curves given above.

The maximal value of the resistance to noise, given by $p\simeq 0.86$, would be reasonable for present-day
technologies were it not for the notorious detection loophole. It is far however from the corresponding value
against a standard quantum eavesdropper, which is around $p\simeq 0.75$. Inside the secure region, however, allowing an eavesdropper limited by no-signalling only does not significantly modify the efficiency of quantum key distribution.

\ack
We acknowledge support by the European Commission under the Integrated Project Qubit Applications (QAP) funded by the IST directorate as Contract Number 015848. AA acknowledges financial support from the Spanish MEC, under a ``Ramon y Cajal" grant. SM acknowledges support by the Interuniversity Attraction Poles
Programme - Belgium Science Policy - under grant V-18 \vfill

\noindent\emph{Note added}. As mentioned earlier we restricted the security analysis of our protocols to individual attacks, in which the eavesdropper tries to acquire information about Alice and Bob's individual systems independently and always in the same way. After this work was submitted, appeared a proof that cryptographic schemes of the type analyzed here are also secure against general attacks by a no-signalling eavesdropper for the more restrictive task of key expansion \cite{mw}. The security proof of \cite{mw} should also apply with small modifications to the protocols presented here. In particular, as mentioned in \cite{mw}, the key rate of one secret bit per ebit in the absence of noise obtained in the present paper also holds in the general framework of \cite{mw}. The security analysis of our protocols against the most general strategies available to a no-signalling eavesdropper remains a problem for future investigations. In general, we expect that stronger security requirements will yield smaller key rates, but that the features of enhanced noise resistance and key rates reported here will continue to hold.

\vspace{8cm}

\end{document}